\documentclass[prl,twocolumn,showpacs,amsmath,amssymb,superscriptaddress]{revtex4}
\usepackage{graphicx}
\usepackage{graphics}
\usepackage{dcolumn}
\usepackage{bm}

\begin{document}
\title{Spin-Hall and spin-diagonal conductivity in the presence of Rashba and Dresselhaus spin-orbit
coupling}
\author{N. A. Sinitsyn}
\affiliation{Department of Physics, Texas A\&M University, College Station, TX 77843-4242}
\author{E. M. Hankiewicz}
\affiliation{Department of Physics, Texas A\&M University, College Station, TX 77843-4242}
\author{Winfried Teizer}
\affiliation{Department of Physics, Texas A\&M University, College Station, TX 77843-4242}
\author{Jairo Sinova}
\affiliation{Department of Physics, Texas A\&M University, College Station, TX 77843-4242}
\date{\today}
\begin{abstract}
We investigate the spin-current linear response conductivity 
tensor to an electric field in a paramagnetic two dimensonal electron gas with
both Rashba and Dresselhaus spin-orbit coupling in the weak scattering regime. 
In the ususal case where both spin-orbit-split bands are occupied, we find that 
the spin-Hall conductivity depends only on the sign of the difference in magnitude of the
Rashba and Dresselhaus coupling strengths except within a narrow window where both
coupling strengths are equal. We also find a new effect in which
a spin current is generated in the direction of the 
driving field whenever the Dresselhaus spin-orbit coupling is nonzero. 
We discuss experimental implications of this finding taking into account the finite mobility 
and typical parameters of current samples and possible experimental set ups for its detection.   
\end{abstract}


\maketitle 

The manipulation of spin by electrical means in semiconducting
enviroments has generated a lot of recent theoretical and experimental research 
aimed to develop useful spintonic devices and novel physical concepts \cite{wolf},
many focusing on effects that generate spin-polarized current \cite{spincurrent}. 
Given the success of ferromagnetic metal based spintronic devices   
\cite{gmr} which have revolutionized the information storage industry, the 
possibility of doping, gating, and heterojunction formation in seminconducting
spintronic devices makes their possibilites that much wider.
However, the practical implementation of semiconducting spintronics is awaiting the 
resolution of effective  injection of spin-polarized carriers 
\cite{spininject} from 
ferromagnetic metals combined with long spin lifetimes \cite{awschalom},
or room-temperature semiconductor ferromagnetism \cite{ohno}.
The recent discovery of the intrinsic spin-Hall effect by Murakami {\it et al} \cite{murakami} in p-doped 
semiconductors and by Sinova {\it et al} \cite{sinova} in Rashba spin-orbit coupled
two dimensional electron gases (2EDGs) offers new avenues in spintronics research and transport
phenomena which may meet the first challenge. 

The intrinsic spin-Hall effect consist 
of a dissipationless spin-current generated perpendicular to the driving electric field in
the weak-scattering limit where spin-orbit coupling is strong. 
This effect contrasts with the extrinsic spin-Hall effect recently revived by Hirsch \cite{hirsch} 
and Zhang \cite{zhang} and first studied by Dyakonov and Perel \cite{dyakonov}, where spin-orbit dependent 
scattering from impurities can generate a Hall spin-current but will vanish in the weak 
scattering limit, where the intrinsic effect dominates.  
The intrinsic spin-Hall effect predicted by Murakami {\it et al} \cite{murakami,murakami2} 
and Sinova {\it et al} \cite{sinova,dimi} has generated interest within the theoretical
community \cite{hu,schliemann_spin_Hall}, supporting the need for a strong 
experimental effort in detecting such an effect.
In the Rashba spin-orbit coupled 2DEGs it was shown that the intrinsic spin-Hall conductivity
has a universal value $e/8\pi$ in the case of both spin-split subbands being occupied \cite{note,sinova}. 
Motivated by recent experiments 
\cite{miller,ganichev2} which have demonstrated the ability to tune the  
magnitude of the Rashba and Dresselhaus spin-orbit coupling strength
directly and by the fact that when both coupling strengths are equal 
the quasiparticles are effectively spin-orbit decoupled \cite{schliemannprl} 
and the intrinsic spin-Hall effect should vanish,  
we extend our prior studies to include 2DEGs with both Rashba and Dresselhaus
spin-orbit coupling. We find that the spin Hall conductivity remains universal in the weak scattering limit
except for a small window where the spin-orbit coupling strengths are equal and changes sign
when the difference of the spin-orbit coupling strenghts change sign. 
We also find a novel effect
in which a spin-polarized current is generated in the direction of the driving electric field
whenever the Dresselhaus spin-orbit coupling is nonzero which should have important consequences if observed.

The Hamiltonian which we consider contains the spin-orbit coupling interaction form for
a two-dimensional electron gas \cite{mishchenko}
\begin{equation}
H_{so}=\lambda (\hat{\sigma}_x p_y-\hat{\sigma}_y p_x)+\beta(\hat{\sigma}_x p_x-
\hat{\sigma}_y p_y),
\end{equation}
where the first term is the Bychkov-Rashba term due to the lack of inversion symmetry
of the trapping well \cite{rashba} and the second is the linear 
Dresselhaus term due to the lack of inversion symmetry in bulk semiconductors \cite{dresselhaus}.
Hence the full particle Hamiltonian can be written as
\begin{equation}
H=\left(\frac{\hat{p}^2}{2m^*}-\mu\right)\sigma_0+\alpha_{ik}\sigma_i p_k,\,\,
\alpha_{ik}=\left(
\begin{array}{cc}
\beta&\lambda\\-\lambda&-\beta
\end{array}\right)
\label{ham}
\end{equation}
This Hamiltonian has a simple spectra given by
\begin{equation}
\xi_\pm({\bf k})=\frac{\hbar^2k^2}{2m^*}-\mu\pm k\sqrt{(\lambda^2+\beta^2)+2\lambda\beta sin(2\phi)}
\label{eval}
\end{equation}
where $\tan\phi=k_y/k_x$, $\mu$ is the chemical potential,
and the corresponding eigenfunctions are given by
\begin{equation}
\psi_\pm({\bf r})=\frac{1}{\sqrt{2}}\left(
\begin{array}{c}
e^{i\theta_k/2}\\\pm e^{-i\theta_k/2}
\end{array}\right)e^{i{\bf k}\cdot{\bf r}}
\label{evec}
\end{equation}
where  
\begin{equation}
\tan{\theta_k}=\frac{\lambda k_x+\beta k_y}{\lambda k_y +\beta k_x}.
\end{equation}

The Rashba coupling  strength in a 2DEG can be modified by a gate field by up to 
50\% \cite{nitta,miller} and therefore this system affords the study to the intricate 
interplay between both types of coupling directly whose ratio $\lambda/\beta$
can vary between 1.5 and 2.5 typically \cite{ganichev2}. Recent observations of a  spin-galvaic 
effect \cite{ganichev,ganichev2} and a spin-orbit coupling weak localization studies
in these systems \cite{miller,peter,koga} illustrate the potential importance of these
tunable interactions in semiconductor spintronics \cite{inoue}. 
Indeed, this intricate interplay has generated several theoretical  studies of their
transport properties due to the anisotropies created in the Fermi surfaces for
$\alpha\ne \beta$ \cite{schliemannprb,mishchenko} and a new proposed spin-FET
\cite{schliemannprl} in the regime where $\lambda=\beta$
motivated by the original proposal by Datta and Das \cite{datta}.

In the Rashba coupled 2DEG ($\beta=0$), Sinova {\it et al} \cite{sinova} found that the dc
z-component spin-current Hall response to a driving internal electric field,
$j_s^z=\sigma^{\rm spin}_{xy} E_y$, within the weak scattering limit has a 
universal value whenever the two Rashba split bands are occupied (the usual case), 
$\sigma^{\rm spin}_{xy}=\frac{e}{8\pi}$, and vanishes linearly with the electron
density when only  one Rashba split band is occupied. This result was obtained both
within the Kubo linear response formalism in the dc-weak scattering limit \cite{sinova}
and within an equivalent and more physically transparent multi-band wave-packet dynamics 
theory \cite{dimi} generalizing the theory introduced by Qian {\it et al} \cite{sundaram}
to account more rigorously for the consequences of the time-dependent effective magnetic
field that is experienced by the electron spins as they move through momentum space.

\begin{figure}
\includegraphics[width=3.in]{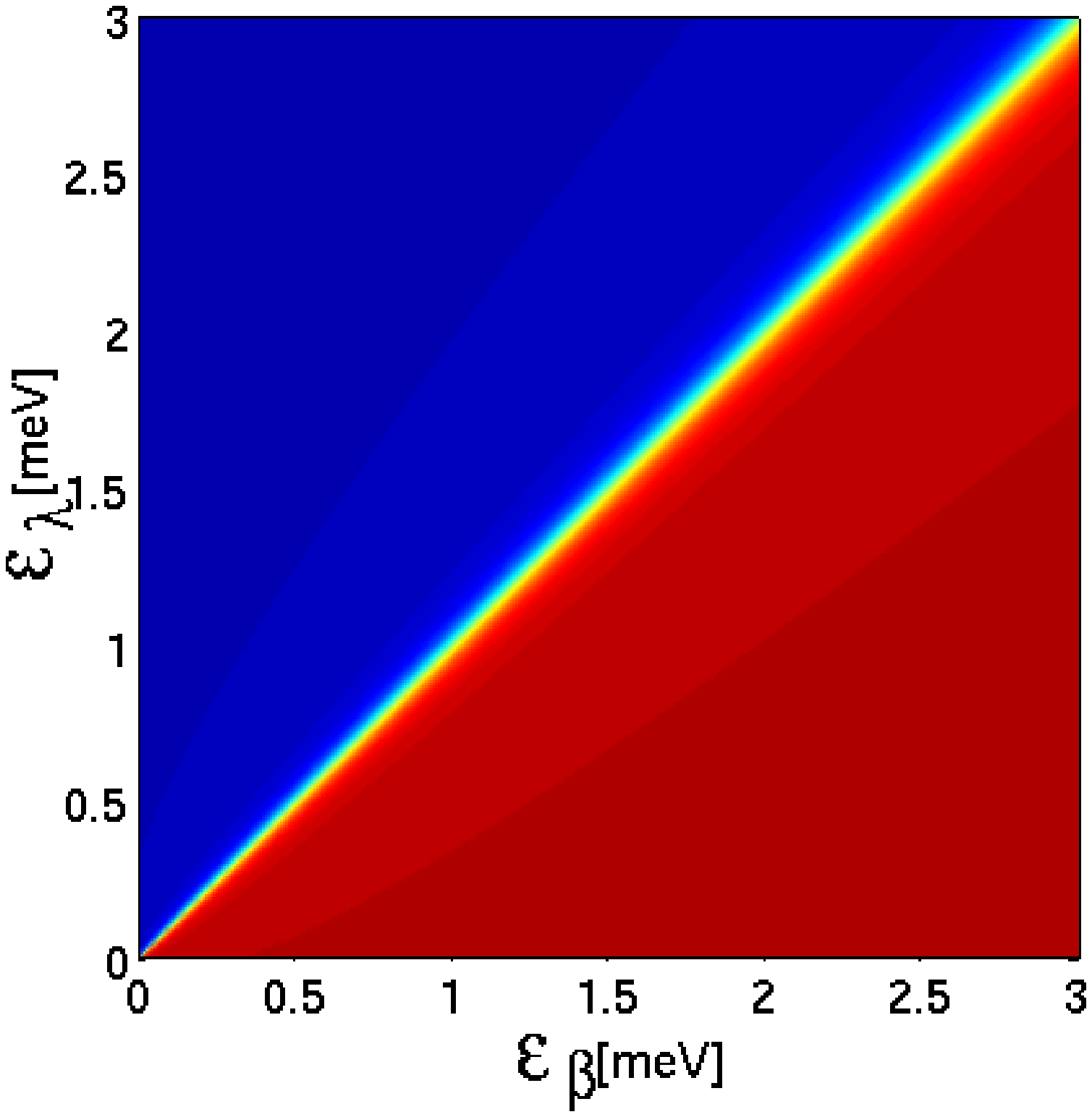}
\includegraphics[width=3.in]{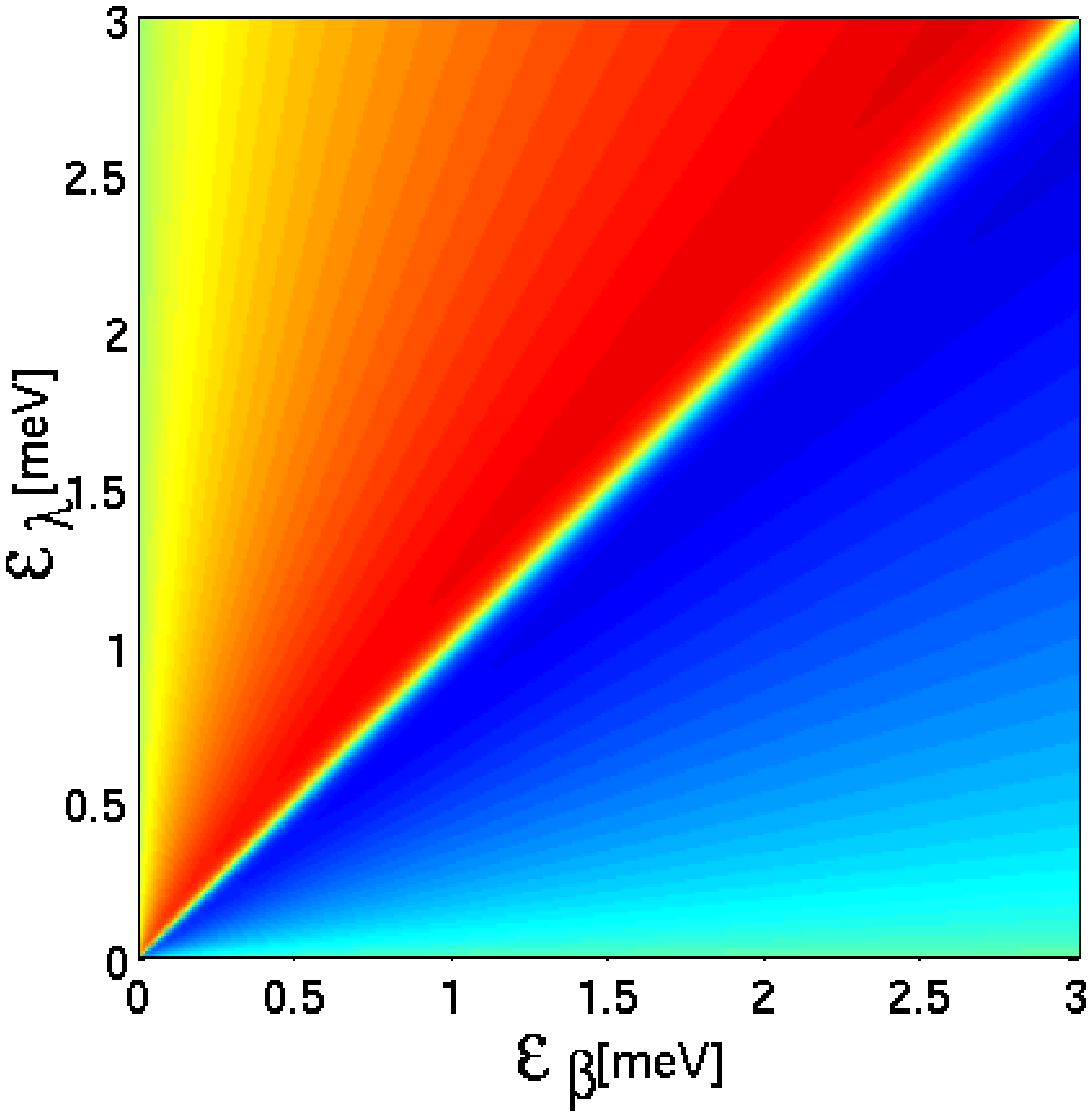}
\caption{Color plot of $\sigma_{xy}^{\rm spin}$ (upper figure) and
$\sigma_{yy}^{\rm spin}$ vs. $\epsilon_\beta=m^*\beta^2/\hbar^2$ and
$\epsilon_\lambda=m^*\beta^2/\hbar^2$ with Fermi energy $\epsilon_F=50$ meV,
$\eta=0.63 meV$, and $m^*/m_e=0.05$.
The color scale is bright red =$e/8\pi$ and dark blue= $-e/8\pi$.}
\label{fig:one}
\end{figure}

To address the interplay between the Rashba and Dresselhaus spin-orbit coupling and its
consequence on the spin-current response to an applied electric field 
we use the linear response Kubo formalism in the weak scattering limit, where as usual
it is assumed that the dc limit is taken with $\omega/\eta>>1$ while
both $\omega$ and $\eta$, the inverse of the quasiparticle lifetime induced by weak scattering,
go to zero \cite{AM}. We will address the effects of finite quasiparticle lifetime,
i.e. finite mobility samples, within the Born approximation 
below similarly to references \cite{sinovaprb} and \cite{schliemann_spin_Hall}.
In this Kubo formalism approach, the intrinsic spin-Hall conductivity
arises as a reactive term in the spin-current response to an electric field
which has the form 
\begin{eqnarray}
\sigma^{\rm spin }_{\alpha y}&=&{ e \hbar} \sum_{n\ne n'}\int \frac{d\vec{k}}{(2\pi)^2}
(f_{n',k}-f_{n,k})\nonumber\\ &\times&
\frac{{\rm Im}[\langle n' k|
\hat{j}^z_{{\rm spin\,\,}\alpha}|nk\rangle\langle nk| \hat{v}_y|n' k\rangle]}
{(\xi_{nk}-\xi_{n'k})^2}
\label{SH}
\end{eqnarray}
where $n,n'=\pm$, $\vec{j}^z_{\rm spin}=\frac{\hbar}{4}\{\hat{\sigma}_z,\vec{v}\}$,
$\vec{v}=\partial{H(k)}/{\partial \hbar \vec{ k}}$ , and $\alpha=x,y$ \cite{inoue}.
From the strong transport anisotropies induced by the change in Fermi surfaces whenever
$\lambda$ and $\beta$ differ, it was expected that the universal spin-Hall conducitivity
found in \cite{sinova} in the weak scattering regime would be significantly modified. Inserting
eigenvalues, Eq. \ref{eval}, and eigenvectors, Eq. \ref{evec}, above we obtain 
(for the experimentally relevant regime of both subbands being occupied):
\begin{eqnarray}
\sigma _{xy}^{z}&=&\left\{
\begin{array}{l}
\frac{e}{8\pi }\,\,\,{\rm for}\,\, \lambda^2>\beta ^{2}\\
\,\,\,0\,\,\,\,\,{\rm for} \,\,\lambda=\beta\\
-\frac{e}{8\pi }\,\,\,{\rm for}\,\, \lambda^2<\beta ^{2}
\end{array}  \right.\\ &&\nonumber\\ 
\sigma _{yy}^{z}&=&\left \{
\begin{array}{l}
-\frac{e}{8\pi }\frac{\beta}{\lambda} \,\,\,{\rm for}\,\, \lambda^2>\beta ^{2}\\
\,\,\,0\,\,\,\,{\rm for}\,\,\lambda=\beta\\
\frac{e}{8\pi }\frac{\lambda}{\beta} \,\,\,{\rm for}\,\, \lambda^2<\beta ^{2}
\end{array}  \right.
\label{pure1}
\end{eqnarray}
These equations predict that the spin-Hall conductivity only depends
on the sign of $\beta^2-\lambda^2$, and that there is a diagonal 
(in the sense that it points along the driving field direction) reactive part of the 
spin-current whenever $\beta$ is non-zero and $\beta\ne \lambda$.
The sharpness and singularities of these results are simply an artifact of the
$\eta\rightarrow 0$ limit which can be easily rectified by introducing a finite
lifetime to the spin-orbit coupled quasiparticles induced by the weak scattering
in the usual Born approximation approach \cite{sinovaprb,mahan} given by
\begin{widetext}
\begin{eqnarray}
\sigma^{spin}_{\alpha y}=-\frac{e}{\omega}\int \frac{d\vec{k} d\epsilon }{(2\pi)^3}
\sum_{n n'} {\rm Im}[\langle n'\vec{k}|\hat{j}^z_{\rm spin\,\,\alpha}|n\vec{k}\rangle
\langle n\vec{k}| \hat{v}_y|n'\vec{k}\rangle] 
f(\epsilon) \left [A_{n',\vec{k}}(\epsilon) {\rm Re}
[G^{\rm ret}_{n,k}(\epsilon+\hbar\omega)]+A_{n,\vec{k}}(\epsilon)
{\rm Re}[G^{\rm adv}_{n',k}(\epsilon-\hbar\omega)]\right],
\label{sigdis}
\end{eqnarray}
\end{widetext}
where  $A_{n,\vec{k}}(\epsilon)=
\eta/((\epsilon-\xi_{n\vec{k}})^2+\eta^2/4)$
is  the disorder  broadened spectral  function and  $G^{\rm  ret/adv}_{n,k}(\hbar\omega)=
1/(\hbar\omega-\xi \pm i \eta/2)$ are  the  advanced and  retarded quasiparticle  Green's
functions  with finite  lifetime  $2\eta^{-1}/\hbar$ (here chosen to be
momentum independent for simplicity). The above expression can be evaluated numerically,
however, we can make progress understanding qualitative the effect of finite quasiparticle 
lifetime by taking the further approximation that the spectral function is sharp enough
to allow one to substitute it for a delta function. This simplification translates into
adding a small complex value to the frequency in the frequency dependent response function
from which the weak scattering dc limit expression, Eq. \ref{SH} was obtained. The expression
that is obtained is identical to Eq. \ref{SH} but with the denominator replaced
by $\eta^2+(\xi_{n\vec{k}}+\xi_{n'\vec{k}})^2$.
Anticipating already the typical experimental situation where the Fermi energy is only 
slightly modified from its non-spin-orbit coupled form, we obtain after some
straightforward algebraic manipulation the expressions
\begin{widetext}
\begin{eqnarray}
\sigma _{xy}^{\rm spin}
=\frac{e}{8\pi}\frac{ \epsilon_{F}(\epsilon_\lambda-\epsilon_\beta)}{\sqrt{ 
\epsilon_{F}^{2}(\epsilon_\lambda-\epsilon_\beta)^2+\eta^2 {\epsilon_F}(\epsilon_\beta+
\epsilon_\lambda)/4+\eta^4/64}}
\label{sxy}
\end{eqnarray}
and
\begin{eqnarray}
\sigma _{yy}^{\rm spin}
= -\frac{e \,(\epsilon_\beta-\epsilon_\lambda)}{16\pi \sqrt{\epsilon_\beta \epsilon_\lambda} }
\left( 1-\frac{\eta
^{2}+8\epsilon_{F}(\epsilon_\lambda+\epsilon_\beta)}{\sqrt{
\eta ^{4}+64\epsilon_{F}^{2}(\epsilon_\beta-\epsilon_\lambda)^{2}+16\eta ^{2}\epsilon_{F}(\epsilon_\beta+\epsilon_\lambda)}}\right)%
\label{syy}
\end{eqnarray}
\end{widetext}
where $\epsilon_F$ is the Fermi energy, and $\epsilon_\lambda= m \lambda^2/\hbar^2$ and
$\epsilon_\beta= m\beta^2/\hbar^2$ are the spin-orbit coupling characteristic energy scales
for the Rashba and Dresselhauss mechanisms defined in the same way as Ref. \cite{schliemann_spin_Hall},
and $\epsilon_F >> \epsilon_\lambda,\epsilon_\beta$ is assumed.
In the limit of $\eta \rightarrow 0$ one recovers Eqs. \ref{sxy} and \ref{syy}, although near the 
$\alpha=\beta$ point we see that one limit goes smoothly but fairly rapidly to the next in the scale of $\eta$
as illustrated in Fig. \ref{fig:one}.
This analysis illustrates our earlier statement \cite{sinova}
indicating that, unlike the universal Hall conductivity value on a 2DEG
quantum Hall plateau, the universality of the intrinsic spin Hall effect does not
extend to the disorder domain.  And, as expected \cite{sinova}
the spin Hall conductivity $\sigma_{xy}^{\rm spin}$ and the spin diagonal
conductivity $\sigma_{yy}^{\rm spin}$ is slightly suppressed for $\epsilon_F,|\epsilon_{\lambda}
-\epsilon_\beta| > \eta$
\begin{eqnarray}
\sigma_{xy}^{\rm spin}&=&\frac{e}{8\pi}\left(1-\frac{\eta^2 (\epsilon_\lambda+\epsilon_\beta)}{8\epsilon_F(\epsilon_\lambda-\epsilon_\beta)^2}\right)
\label{sigxydis}\\
\sigma_{yy}^{\rm spin}&=&\frac{e}{8\pi}\sqrt{\frac{\epsilon_\beta}{\epsilon_\lambda}}
\left(1-\frac{ \eta^2 \epsilon_{\lambda}}{4 \epsilon_F(\epsilon_\lambda-\epsilon_\beta)^2}\right)
\label{sigyydis}
\end{eqnarray}
for $\epsilon_\lambda>\epsilon_\beta$, and for  $\epsilon_\lambda<\epsilon_\beta$
the equations have opposite sign and we must switch
 $\epsilon_\beta$ and $\epsilon_\lambda$ in Eq. \ref{sigyydis}. We note that 
Eq. \ref{sigxydis} is in agreement with Ref. \cite{schliemann_spin_Hall} for $\beta=0$. 
The strong disorder limit
of Eqs. \ref{sxy} and \ref{syy} the spin conductivity tensor goes as 
$\epsilon_F\epsilon_{\lambda/\beta}/\eta^2$ but in this case one must still assume 
$\epsilon_F/\eta>>1$ (a basic justification of the above approximation) and therefore
$\epsilon_{\lambda/\beta}/\eta<<1$ is the condition for such effects to vanish. In such
a limit we expect that the extrinsic contribution from skew scattering 
proposed by Hirsch may dominate since at this stage scattering from impurities may be
more important than the intrinsic spin-orbit coupling subband splitting. 
We also note that the specific coefficients in the above expression proportional to $\eta^2$ 
will change since the full expression \ref{sigdis} which takes into account the 
broadening present in the occupation number must be used in a quantitative analysis.

In current 2DEGs high quality samples \cite{nitta,grundler,miller,ganichev2} the typical carrier
concentrations
range from $5\times 10^{11}$ to $10^{12}$ cm$^{-2}$, the $\lambda,\beta$ range is $1-5 \times 10^{-11}$ 
eVm, and the mobilities range is $1-50$ m$^2$/Vs. In terms of the effective mass ratio $m^*/m_e$
the energy scales defined above are given by
\begin{eqnarray}
\epsilon_F&\approx& 0.24 \frac{n_{2D}[10^{11}{\rm cm}^{-2}]}{m^*/m_e},\\ 
\,\,\epsilon_{\lambda/\beta}&\approx& 1.31 \frac{m^*}{m_e}(\lambda/\beta[10^{-11} eVm])^2\\
\eta&\approx& \frac{0.116}{\mu[m/Vs] (m^*/m_e)},
\end{eqnarray}
with $\mu$ being in this case the mobility of the sample. For
 the typical $m^*/m_e=0.05$ in InAs based heterostructures gives a range of
$\epsilon_F=20-50$ meV, $\epsilon_{\lambda/\beta}=0.07-1.6$ meV, and $\eta=0.05-2.3 meV$. 
This indicates that current samples (although not all) are already within the regime
where these effects should be observable and the weak scattering regime applicable.
Outside this weak scattering regime we note that will be a further
contribution to $\sigma_{sH}$, like that envisaged by Hirsch \cite{hirsch},
but in the limit emphasized here (weak scattering limit)  it will not dominate
in the high-mobility 2DEG's \cite{note2}.

It is also important to note that 
the intrinsic character of our spin Hall effect, compared to the extrinsic
character of the effect discussed by Hirsch \cite{hirsch}, is
analogous to the intrinsic contribution to the anomalous Hall
effect recently emphasize in various studies \cite{sundaram,jungwirth,
nagaosa1,sinovaprb,culcer},  proposed to be the main contribution
to the anomalous Hall effect in some ferromagnets and strongly polarized paramagnets.  
In both cases scattering contributions to the Hall conductivities can become important if skew
scattering  \cite{bruno} is present, especially when the overall electron
scattering rate is small and the steady state distribution function of the
current-carrying state is strongly disturbed compared to the equilibrium one.
In the case of (III,Mn)V ferromagnetic semiconductors for example,
the intrinsic theory of the anomalous Hall effect accounts rather 
convincingly for experimental observations \cite{jungwirth}.

Besides the several approaches already proposed to measure the spin-Hall effect
\cite{murakami,sinova,culcer,murakami2}, the tunability of both the Rashba and Dresselhaus
coupling parameters \cite{ganichev2,miller} and the non-zero diagonal spin-current generated
in the presence of a linear Dresselhaus term offers new possibilities. We propose a bottom
gate sample coupled with non-contact probes such as spatially resolved Kerr effect measuremnts,
ferromagnetic STM, or Scanning Hall probe microscope. Having four non-contact local probe observations 
(two for Hall and two for diagonal conductivity) on the edges of the sample as a function
of gate voltage which varies the Rashba and Dresselhaus coupling \cite{ganichev2,miller}
simultaneous acquisition of transverse and diagonal spin currents without the usual spurious geometrical
effects present in typical transport measurements.

\begin{acknowledgments}
The authors would like to thank T. Jungwirth, A. H. MacDonald, S. Murakami, N. Nagaosa, Q. Niu,
and S.-C. Zhang for extensive and fruitfuil discussions, and J. Schliemann and D. Loss for
useful correspondence. N. A. S. was supported by DOE under grant No. DE-FG03-96ER45598,
NSF under grant DMR0072115
and Telecommunication and Information Task Force at TAMU.
\end{acknowledgments}


\begin{references}
\bibitem{note} The word universality is here meant 
in the sense that it does not depend on material parameters and it is given only in terms
of universal constants.
\end{references}
\end{document}